\documentclass[11pt,twoside]{article}
\usepackage{epsfig}

\usepackage{amsmath}
\usepackage{amssymb}
\usepackage{cite}

\begin{document}
\newcommand{\pst}{\hspace*{1.5em}}

\newcommand{\rigmark}{\em Journal of Russian Laser Research}
\newcommand{\lemark}{\em Volume 37, Number 1, 2016}

%\lhead[\fancyplain{\rigmark, {\em \lemark}}{\rigmark}]{\fancyplain{\rigmark, {\em \lemark}}{\lemark}}
%\chead{}\rhead[\fancyplain{}{\lemark}]{\fancyplain{}{\rigmark}}
%\plainfootrulewidth 0.4pt
\newcommand{\be}{\begin{equation}}
\newcommand{\ee}{\end{equation}}
\newcommand{\bm}{\boldmath}
\newcommand{\ds}{\displaystyle}
\newcommand{\bea}{\begin{eqnarray}}
\newcommand{\eea}{\end{eqnarray}}
\newcommand{\ba}{\begin{array}}
\newcommand{\ea}{\end{array}}
\newcommand{\arcsinh}{\mathop{\rm arcsinh}\nolimits}
\newcommand{\arctanh}{\mathop{\rm arctanh}\nolimits}
\newcommand{\bc}{\begin{center}}
\newcommand{\ec}{\end{center}}
\renewcommand{\thefootnote}{\fnsymbol{footnote}}

\thispagestyle{plain}
\label{sh}

%\lfoot[\fancyplain{\ \\[1mm] \thepage}{\ \\[1mm]\thepage}]{\fancyplain{}{}}

\begin{center} {\Large \bf
%\begin{tabular}{c}
HIDDEN BELL CORRELATIONS\\[1mm]
IN THE FOUR-LEVEL ATOM\footnotemark{}
\footnotetext{Based on
% ~~~~~~~ A fragment of
the invited talk delivered by M.~A.~Man'ko at the 46th Winter Colloquium on
the Physics of Quantum Electronics (PQE-2016, January 3--8, 2016, Snowbird,
Utah, USA). }
%\\[-1mm]
%\end{tabular}
 } \end{center}

\bigskip

\begin{center}

{\bf Margarita A. Man'ko$^{1\,*}$ and Vladimir I. Man'ko$^{1,\,2}$
}\end{center}

\medskip

\begin{center}
{\it $^1$Lebedev Physical Institute, Russian Academy of Sciences\\ Leninskii
Prospect 53, Moscow 119991, Russia}

\smallskip

{\it $^2$Moscow Institute of Physics and Technology (State University)\\
Dolgoprudnyi, Moscow Region 141700, Russia}

\medskip
$^*$Corresponding author email:~~~mmanko@sci.lebedev.ru\\

\end{center}

\begin{abstract}\noindent
We extend the Bell inequality known for two qubits to the four-level atom,
including an artificial atom realized by the superconducting circuit, and
qudit with $j=3/2$. We formulate the extended inequality as the inequality
valid for an arbitrary Hermitian nonnegative 4$\times$4-matrix with unit trace
for both separable and entangled matrices.
\end{abstract}

\medskip


\section{Introduction}
\pst Bell inequalities~\cite{Bell,CHSH} are connected with properties of
quantum correlations. The quantum correlations in composite systems
containing subsystems correspond to the influence of the behavior of
the subsystem degrees of freedom on the behavior of the degrees of
freedom of the other subsystems.

For bipartite systems, there exist entropic inequalities like the
subadditivity condition providing the nonnegativity of mutual von
Neumann information. This inequality is also connected with quantum
correlations of the subsystem degrees of freedom in the composite
system.

For tripartite systems, the strong subadditivity
condition~\cite{Lieb-Ruskai} characterizes the level of quantum
correlations of the subsystem degrees of freedom in the system.
Recently, it was pointed out~\cite{my1,my2,my3,509MAM} that
analogous inequalities exist in noncomposite systems as well. This
fact was understood due to the application of a specific invertible
map of integers $s=1,2,\ldots,N$, where $N=nm$, onto pairs of
integers $(jk)$, where $j=1,2,\ldots n$ and $k=1,2,\ldots m$, or, in
the case $N=n_1n_2n_3$, to triples of integers $(jkl)$, where
$j=1,2,\ldots n_1$, $k=1,2,\ldots n_2$, and $l=1,2,\ldots n_3$.

The aim of this work is to extend the Bell
inequalities~\cite{Bell,CHSH} known for bipartite system states of
two qubits to the noncomposite system states like the four-level
atom or qudit with $j=3/2$. We show that the
Clauser--Horne--Shimony--Holt~(CHSH) inequality~\cite{CHSH} can be
rewritten for an arbitrary Hermitian 4$\times$4-matrix $\rho\geq 0$,
such that Tr$~\rho=1$. In the case of qudit ($j=1$) states, the
violation of the Bell inequalities corresponds to quantum
correlations, which we call the hidden Bell correlations.

This paper is organized as follows.

In Sec.~2, we consider the system of two qubits and recently
obtained entropic inequalities like the subadditivity condition for
a single qudit with $j=3/2$. In Sec.~3, we review analogous new
subadditivity condition and Araki--Lieb inequalities for qutrit
states. In Sec.~4, we discuss tomographic probability distributions
of spin
states~\cite{OlgaJETP,DodPLA,ScullyFPL,NuovoCimento-Benasque}. In
Sec.~5, we write the Bell inequality for a single qudit and the
four-level atom and present the new inequality (an analog of the
subadditivity condition) for an arbitrary Hermitian nonnegative
matrix with unit trace in Sec.~6. In the concluding Sec.~7, we give
the prospective of the approach elaborated, including the
application to the superconducting
circuit~\cite{OlgaDodon1990,ZeilingerJPA,ZeilingerPRB,Kiktenko1,Kiktenko2,GlushkovJRLR}.
In the Appendix, we provide the readers with an analog of the new
Bell inequality for an arbitrary Hermitian 4$\times$4-matrix.

\section{Quantum System of Two Qubits}

\pst If we use the map of indices\newline
$~\qquad\qquad\qquad\qquad\qquad 1/2~1/2\leftrightarrow 1,~
1/2\,-1/2\leftrightarrow 2,~-1/2~1/2\leftrightarrow
3,~-1/2\,-1/2\leftrightarrow 4$,\newline the density matrix
$\rho_{m_1m_2m'_1m'_2}$ of two qubits with $m_1,m_2,m'_1,m'_2=\pm
1/2$ can be written explicitly in two forms

\vspace{5mm}

\noindent \scalebox{0.92}{$
\rho=\left(\begin{array}{cccc}
\rho_{11}&\rho_{12}&\rho_{13}&\rho_{14}\\
\rho_{21}&\rho_{22}&\rho_{23}&\rho_{24}\\
\rho_{31}&\rho_{32}&\rho_{33}&\rho_{34}\\
\rho_{41}&\rho_{42}&\rho_{43}&\rho_{44}
\end{array}\right)\equiv\left(\begin{array}{cccc}
\rho_{1/2\,1/2\,1/2\,1/2}&\rho_{1/2\,1/2\,1/2\,-1/2}&\rho_{1/2\,1/2\,-1/2\,1/2}&\rho_{1/2\,1/2\,-1/2\,-1/2}\\
\rho_{1/2\,-1/2\,1/2\,1/2}&\rho_{1/2\,-1/2\,1/2\,-1/2}&\rho_{1/2\,-1/2\,-1/2\,1/2}&\rho_{1/2\,-1/2\,-1/2\,-1/2}\\
\rho_{-1/2\,1/2\,1/2\,1/2}&\rho_{-1/2\,1/2\,1/2\,-1/2}&\rho_{-1/2\,1/2\,-1/2\,1/2}&\rho_{-1/2\,1/2\,-1/2\,-1/2}\\
\rho_{-1/2\,-1/2\,1/2\,1/2}&\rho_{-1/2\,-1/2\,1/2\,-1/2}&\rho_{-1/2\,-1/2\,-1/2\,1/2}&\rho_{-1/2\,-1/2\,-1/2\,-1/2}\end{array}
        \right).$}\\[5mm]
We denote the matrix $\rho\equiv\rho(1,2)$. The density matrix
$\rho_{ss'}~(s,s'=1,2,3,4)$ rewritten in the form
$\rho_{m_1m_2m'_1m'_2}$ is Hermitian, $\rho^\dagger=\rho$, with
Tr$\,\rho=1$, and it is nonnegative, $\rho\geq 0$, which means that
its eigenvalues $\lambda_1,\lambda_2,\lambda_3,\lambda_4$ are
nonnegative numbers.

For two qubit systems, one has the density matrices for each qubit
state obtained by the procedure of partial tracing. The density
matrices of the first qubit $\rho(1)$ with matrix elements
$\rho(1)_{m_1m'_1}$ and the second qubit $\rho(2)$ with matrix
elements $\rho(2)_{m_2m'_2}$, by definition, read
\begin{eqnarray*}
\rho(1)_{m_1\,m'_1}=\big(\mbox{Tr}_2\rho(1,2)\big)_{m_1\,m'_1}=
\sum_{m_2=-1/2}^{1/2}\big(\rho(1,2)\big)_{m_1\,m_2\,m'_1\,m_2}~,\\
\rho(2)_{m_2\,m'_2}=\big(\mbox{Tr}_1\rho(1,2)\big)_{m_2\,m'_2}=
\sum_{m_1=-1/2}^{1/2}\big(\rho(1,2)\big)_{m_1\,m_2\,m_1\,m'_2}~.
\end{eqnarray*}
One can check that, following these definitions, we obtain
explicitly matrices $\rho(1)$ and $\rho(2)$ in terms of $\rho_{ik}$
in two forms:
$$\rho(1)=\left(\begin{array}{cc}
\rho(1)_{1/2~1/2}&\rho(1)_{1/2~-1/2}\\
\rho(1)_{-1/2\,~1/2}&\rho(1)_{-1/2~-1/2}\end{array}\right)\quad\mbox{or}\quad
\rho(1)=\left(\begin{array}{cc}
\rho_{11}+\rho_{22}&\rho_{13}+\rho_{24}\\
\rho_{31}+\rho_{42}&\rho_{33}+\rho_{44}\end{array}\right),$$
and
$$\rho(2)=\left(\begin{array}{cc}
\rho(2)_{1/2~1/2}&\rho(2)_{1/2~-1/2}\\
\rho(2)_{-1/2\,~1/2}&\rho(2)_{-1/2~-1/2}\end{array}\right)\quad\mbox{or}\quad
\rho(2)=\left(\begin{array}{cc}
\rho_{11}+\rho_{33}&\rho_{12}+\rho_{34}\\
\rho_{21}+\rho_{43}&\rho_{22}+\rho_{44}\end{array}\right).$$

For matrices $~\rho(1,2)$, $~\rho(1)$, and $~\rho(2)$, there exists the
well-known quantum subadditivity condition for von Neumann entropies:
$$S(1,2)=-\mbox{Tr}\,\rho(1,2)\ln\rho(1,2),~~
S(1)=-\mbox{Tr}\,\rho(1)\ln\rho(1),~~
S(2)=-\mbox{Tr}\,\rho(2)\ln\rho(2);$$ it reads $~S(1,2)\leq
S(1)+S(2).$

In the form of matrix inequality written in terms of the matrix
$\rho_{jk}$ ($j,k=1,2,3,4$), the quantum subadditivity condition for
von Neumann entropies is
 $$-\mbox{Tr}\left(\begin{array}{cccc}
\rho_{11}&\rho_{12}&\rho_{13}&\rho_{14}\\
\rho_{21}&\rho_{22}&\rho_{23}&\rho_{24}\\
\rho_{31}&\rho_{32}&\rho_{33}&\rho_{34}\\
\rho_{41}&\rho_{42}&\rho_{43}&\rho_{44}
\end{array}\right)\ln\left(\begin{array}{cccc}
\rho_{11}&\rho_{12}&\rho_{13}&\rho_{14}\\
\rho_{21}&\rho_{22}&\rho_{23}&\rho_{24}\\
\rho_{31}&\rho_{32}&\rho_{33}&\rho_{34}\\
\rho_{41}&\rho_{42}&\rho_{43}&\rho_{44}
\end{array}\right)$$
$$\leq -\mbox{Tr}\left(\begin{array}{cc}
\rho_{11}+\rho_{12}&\rho_{13}+\rho_{24}\\
\rho_{31}+\rho_{42}&\rho_{33}+\rho_{44}
\end{array}\right)\ln\left(\begin{array}{cc}
\rho_{11}+\rho_{12}&\rho_{13}+\rho_{24}\\
\rho_{31}+\rho_{42}&\rho_{33}+\rho_{44}
\end{array}\right)$$
$$-\mbox{Tr}\left(\begin{array}{cc}
\rho_{11}+\rho_{33}&\rho_{12}+\rho_{34}\\
\rho_{21}+\rho_{43}&\rho_{22}+\rho_{44}
\end{array}\right)\ln\left(\begin{array}{cc}
\rho_{11}+\rho_{33}&\rho_{12}+\rho_{34}\\
\rho_{21}+\rho_{43}&\rho_{22}+\rho_{44}
\end{array}\right).$$
The inequality is the matrix inequality for the Hermitian density
matrix $\rho_{jk}$, i.e., $\rho^*_{jk}=\rho_{kj}$ with Tr$\,\rho=1$,
i.e., $\sum_k\rho_{kk}=1$, with nonnegative eigenvalues. The
inequality is valid independently of our quantum-mechanical
interpretation, and it is valid for qudit with $j=3/2$. To see this,
we should use the map\newline
 $~\qquad\qquad\qquad\qquad\qquad 3/2\leftrightarrow
1,\quad 1/2\leftrightarrow 2,\quad -1/2\leftrightarrow 3,\quad
-3/2\leftrightarrow 4$\newline in the above inequality.

\section{Qutrit Density Matrix as a 4$\times$4 Matrix}

\pst Now we present this inequality in the form, which is applicable to the
density matrix of noncomposite qutrit $(j=1)$ state
$\rho_{mm'}=\left(\begin{array}{ccc}
\rho_{1\,1}&\rho_{1\,0}&\rho_{1~-1}\\
\rho_{0\,1}&\rho_{0\,0}&\rho_{0~-1}\\
\rho_{-1\,1}&\rho_{-1\,0}&\rho_{-1~-1}
\end{array}\right),
$ where spin projections $m,m'=1,0,-1$. We use the embedding
$$\rho_{mm'}=\left(\begin{array}{ccc}
\rho_{1\,1}&\rho_{1\,0}&\rho_{1~-1}\\
\rho_{0\,1}&\rho_{0\,0}&\rho_{0~-1}\\
\rho_{-1\,1}&\rho_{-1\,0}&\rho_{-1~-1}
\end{array}\right)\to\rho=\left(\begin{array}{cccc}
\rho_{1\,1}&\rho_{1\,0}&\rho_{1~-1}&0\\
\rho_{0\,1}&\rho_{0\,0}&\rho_{0~-1}&0\\
\rho_{-1\,1}&\rho_{-1\,0}&\rho_{-1~-1}&0\\
0&0&0&0\end{array}\right).$$ We got the 4$\times$4-matrix $\rho$,
which we can consider as the density matrix of ``two artificial
qubits'' already studied above. Also we can apply the inequality
found above for any 4$\times$4-matrix and obtain a new inequality
for the matrix $\rho_{jk}$ $(j,k=1,2,3)$.

The new inequality~\cite{OlgaVova-JRLR2013} for the matrix
$\rho_{jk}$ $(j,k=1,2,3)$ reads
$$-\mbox{Tr}\left(\begin{array}{ccc}
\rho_{11}&\rho_{12}&\rho_{13}\\
\rho_{21}&\rho_{22}&\rho_{23}\\
\rho_{31}&\rho_{32}&\rho_{33}
\end{array}\right)\ln\left(\begin{array}{ccc}
\rho_{11}&\rho_{12}&\rho_{13}\\
\rho_{21}&\rho_{22}&\rho_{23}\\
\rho_{31}&\rho_{32}&\rho_{33}
\end{array}\right)$$
$$\leq -\mbox{Tr}\left(\begin{array}{cc}
\rho_{11}+\rho_{22}&\rho_{13}\\
\rho_{31}&\rho_{33}
\end{array}\right)\ln\left(\begin{array}{cc}
\rho_{11}+\rho_{22}&\rho_{13}\\
\rho_{31}&\rho_{33}
\end{array}\right)
-\mbox{Tr}\left(\begin{array}{cc}
\rho_{11}+\rho_{33}&\rho_{12}\\
\rho_{21}&\rho_{22}
\end{array}\right)\ln\left(\begin{array}{cc}
\rho_{11}+\rho_{33}&\rho_{12}\\
\rho_{21}&\rho_{22}
\end{array}\right).$$
This inequality is valid for an arbitrary Hermitian
3$\times$3-matrix $\rho$ with Tr$\,\rho=1$ and nonnegative
eigenvalues. On the other hand, it is an analog of the new
subadditivity condition for qutrit states, which has no subsystems;
it reflects hidden correlations in the three-level atomic states.

As an example, we present the Araki--Lieb
inequality~\cite{Araki-Lieb}
$$S(1,2)\geq|S(1)-S(2)|$$
(known for von Neumann entropies of bipartite system and its
subsystems) for entropy of the three-level atomic state
(noncomposite system) in the form
$$-\mbox{Tr}\begin{pmatrix}
\rho_{11}&\rho_{12}&\rho_{13}\\
\rho_{21}&\rho_{22}&\rho_{23}\\
\rho_{31}&\rho_{32}&\rho_{33}
\end{pmatrix}\ln\begin{pmatrix}
\rho_{11}&\rho_{12}&\rho_{13}\\
\rho_{21}&\rho_{22}&\rho_{23}\\
\rho_{31}&\rho_{32}&\rho_{33}
\end{pmatrix}$$
$$\geq\left|\mbox{Tr}\begin{pmatrix}
\rho_{11}+\rho_{22}&\rho_{13}\\
\rho_{31}&\rho_{33}
\end{pmatrix}\ln\begin{pmatrix}
\rho_{11}+\rho_{22}&\rho_{13}\\
\rho_{31}&\rho_{33}
\end{pmatrix}-\mbox{Tr}\begin{pmatrix}
\rho_{11}+\rho_{33}&\rho_{12}\\
\rho_{21}&\rho_{22}
\end{pmatrix}\ln\begin{pmatrix}
\rho_{11}+\rho_{33}&\rho_{12}\\
\rho_{21}&\rho_{22}
\end{pmatrix}\right|;$$
this is a new inequality for the qutrit state.

The inequalities can be checked in the experiments where the density
matrices of the qutrit states are measured.

\section{Tomographic Distribution for Spin-$j$ States}
\pst
Quantum states with the density matrix $\rho$ are also determined by
quantum tomograms, which for qudit states are fair probability
distributions defined
as~\cite{OlgaJETP,DodPLA,ScullyFPL,NuovoCimento-Benasque}
$$w(m,u)=\langle m\mid u\rho u^\dagger\mid m\rangle$$ of random spin
projections $m$ depending on the unitary matrix $u$.

The inequality for two quantum tomograms $w_1(m,u)$ and
$w_2(m,u)$ can be written %~\cite{4,5}
in the form of the positivity condition for relative entropy:
$$\sum_{m=-j}^jw_1(m,u)\ln\dfrac{w_1(m,u)}{w_2(m,u)}\geq 0$$ known for
classical probability distributions, but this inequality is written
for quantum systems. The inequality is valid for an arbitrary
unitary $N$$\times$$N$-matrix $u$.

For two qubits, the tomogram of the quantum state with the density
matrix $\rho=\rho(1,2)$ reads
$$
w(m_1,m_2,u_1,u_2)=\langle m_1m_2\mid u_1\times
u_2\,\rho(1,2)\,u_1^\dagger\times u_2^\dagger\mid m_1m_2),$$ where
$m_1,m_2=\pm 1/2$ are spin projections and $u_1,u_2$ are unitary
2$\times$2 matrices of local transforms. The matrices can be labeled
by unit vectors $\vec n_1$ and $\vec n_2$, and this means that the
tomogram is the joint probability distribution $w(m_1,m_2,\vec
n_1,\vec n_2)$ of the spin projection $m_1$ and $m_2$ on the
directions $\vec n_1$ and $\vec n_2$, respectively.

\section{Bell Inequalities for the Four-Level Atom}
\pst

The Bell inequality for separable states of two qubits (two spins
1/2) are given in the form of inequality for correlations in the
system, namely,
$$
|B|=|\langle m_1m_2\rangle_{\vec a\,\vec b}+\langle
m_1m_2\rangle_{\vec a\,\vec c}+ \langle m_1m_2\rangle_{\vec d\,\vec
b}- \langle m_1m_2\rangle_{\vec d\,\vec c}|\leq 2.$$ Here, $\vec a$,
$\vec b$,  $\vec c$, and  $\vec d$ are directions given by unit
vectors orthogonal to the sphere and determined by pairs of Euler
angles $(\varphi_k,\theta_k)$, $k=1,2,3,4$. Here, the numbers $m_1$
and $m_2$ equal to $\pm 1$ are the spin projections multiplied by 2
of the first spin and the second spin, respectively. The projection
of the first spin $m_1$ is measured on the direction given by
vectors $\vec a$ and  $\vec b$. The projection of the second spin
$m_2$ is measured on the direction given by vectors $\vec c$ and
$\vec d$. The number $B$ reads

\vspace{5mm}

\noindent \scalebox{0.92}{$
B=\mbox{Tr}\left[\left(\begin{array}{cccc}
1&-1&-1&1\\
1&-1&-1&1\\
1&-1&-1&1\\
-1&1&1&-1\end{array}\right) \left(\begin{array}{cccc}
w\left(+\frac12,+\frac12,\vec a,\vec b\right)
&w\left(+\frac12,+\frac12,\vec a,\vec c\right)
&w\left(+\frac12,+\frac12,\vec d,\vec b\right)
&w\left(+\frac12,+\frac12,\vec d,\vec c\right) \\
w\left(+\frac12,-\frac12,\vec a,\vec b\right)
&w\left(+\frac12,-\frac12,\vec a,\vec c\right)
&w\left(+\frac12,-\frac12,\vec d,\vec b\right)
&w\left(+\frac12,-\frac12,\vec d,\vec c\right) \\
w\left(-\frac12,+\frac12,\vec a,\vec b\right)
&w\left(-\frac12,+\frac12,\vec a,\vec c\right)
&w\left(-\frac12,+\frac12,\vec d,\vec b\right)
&w\left(-\frac12,+\frac12,\vec d,\vec c\right) \\
w\left(-\frac12,-\frac12,\vec a,\vec b\right)
&w\left(-\frac12,-\frac12,\vec a,\vec c\right)
&w\left(-\frac12,-\frac12,\vec d,\vec b\right)
&w\left(-\frac12,-\frac12,\vec d,\vec c\right)
\end{array}\right)\right],$}
% \par
where the matrix elements in each column are tomographic
probabilities of the states of two spins determined by the
4$\times$4 matrix $\rho(1,2)$.

For the separable state, $|B|<2$.

For entangled state, $|B|\leq 2\sqrt 2$.

We can rewrite the Bell inequality in a matrix form.

We use that the tomographic probabilities $w(m_1,m_2,\vec n_1,\vec
n_2)$ are equal to diagonal elements of a 4$\times$4-matrix, which
is the density matrix transformed by unitary matrix $u_1\otimes
u_2$,~i.e.,
$$
\left(u_1\otimes u_2\cdot \rho(1,2)\cdot u_1^\dagger\otimes
u_2^\dagger\right)_{m_1m_2,m_1m_2}=w(m_1,m_2,\vec n_1,\vec n_2).$$
Here, $u_1$ and $u_2$ are unitary 2$\times$2-matrices of the form
$$u(\varphi,\theta,\psi)=\left(\begin{array}{cc}
\cos\theta/2\exp[i(\varphi+\psi)/2]~&~\sin\theta/2\exp[i(\varphi-\psi)/2]\\
-\sin\theta/2\exp[-i(\varphi-\psi)/2]~&~\cos\theta/2\exp[-i(\varphi+\psi)/2]\end{array}\right).
$$
One can check that the tomogram $w(m_1,m_2,\vec n_1,\vec n_2)$
depends only on pairs of angles determining the directions given by
the vectors $\vec n_1$ and $\vec n_2$.

The presentation of the Bell number in the matrix form and the use
of the map of indices provide the possibility to extend the Bell
inequality to any density 4$\times$4-matrix $\rho$, including the
density matrix of the four-level atom or the density matrix of qudit
with $j=3/2$.

The inequality for these systems is as follows.

Given the nonnegative Hermitian matrix $\rho_{jk}~(j,k=1,2,3,4)$
with $\mbox{Tr}\,\rho=1$, which can be presented in the
separable-state form. We construct the diagonal matrix elements of
the matrix $$\left(u_1\otimes u_2\cdot \rho\cdot u_1^\dagger\otimes
u_2^\dagger\right)_{kk}=\Omega(k|u_1,u_2).$$ The matrix elements are
nonnegative and satisfy the normalization condition
$\sum_{k=1}^4\Omega(k|u_1,u_2)=1.$ The dependence on the unitary
matrices $u_1$ and   $u_2$ is the dependence on angles
$\varphi_1,\theta_1$ and $\varphi_2,\theta_2$.

Then a ``new'' Bell inequality follows from the matrix form of the
CGSH inequality
\begin{eqnarray*}
&&\Big|\Omega(1|u_a,u_b)-\Omega(2|u_a,u_b)-\Omega(3|u_a,u_b)+\Omega(4|u_a,u_b)\\
&&+\Omega(1|u_a,u_c)-\Omega(2|u_a,u_c)-\Omega(3|u_a,u_c)+\Omega(4|u_a,u_c)\\
&&+\Omega(1|u_d,u_b)-\Omega(2|u_d,u_b)-\Omega(3|u_d,u_b)+\Omega(4|u_d,u_b)\\
&&-\Omega(1|u_d,u_c)|+\Omega(2|u_d,u_c)+\Omega(3|u_d,u_c)-\Omega(4|u_d,u_c)\Big|\leq
2.\end{eqnarray*} This inequality is valid for any density
4$\times$4-matrix $\rho$, which can be presented in a separable form
$$\rho=\sum_np_n\rho^{(n)}(1)\otimes\rho^{(n)}(2),\qquad 1\geq
p_n\geq 0,\qquad\sum_np_n=1,$$ where 2$\times$2-matrices
$\rho^{(n)}(1)$ and $\rho^{(n)}(2)$ are nonnegative Hermitian
matrices with unit trace.

The numerical inequality for the 4$\times$4-matrix $\rho$ does not
depend on the interpretation of this matrix as the density matrix.
This inequality can be checked experimentally, e.g., for the
four-level atomic state realized by the superconducting circuit.

For an arbitrary 4$\times$4-matrix $\rho$ with nonnegative
eigenvalues, unit trace, and such that $\rho^\dagger=\rho$, the
bound in this inequality is equal to $2\sqrt 2$; this
Tsirelson\footnote{Concerning Cirelson and Tsirelson, see
[http://www.tau.ac.il/~tsirel/faq1.html]; living in Russia till
1991, Prof.~B.~S.~Cirelson wrote most papers in Russian but after
1991 all his publications are written in English and he uses
B.~Tsirelson.}~\cite{Tsirelson} bound takes place for ``entangled''
matrix $\rho$, for example, $\rho=\dfrac12
\left(\begin{array}{cccc}
1&0&0&1\\
0&0&0&0\\
0&0&0&0\\
1&0&0&1\end{array}\right).$ This matrix corresponds to the pure
state $\mid\psi\rangle$ of qudit with $j=3/2$ of the form $
\mid\psi\rangle=2^{-1/2}\big(\mid 3/2\rangle+\mid -3/2\rangle\big)$.

In this case, the violation of Bell inequality is related to
correlations between the states with different spin projections. An
analogous inequality is valid for the qutrit density 3$\times$3
matrix presented in the form of 4$\times$4 matrix $\rho$ with zero
matrix elements. The same is true for the qubit state.

Thus, we can write formal Bell inequality for single qubits,
qutrits, two qubits, qudit with $j=3/2$, and four-level atom.

The inequality with the Tsirelson bound $2\sqrt 2$ reflects the
strong hidden quantum Bell correlations in these systems.

\section{Matrix Inequality (Subadditivity Condition)}
\pst
Now we construct two positive maps $\rho\to\rho(1)$ and
$\rho\to\rho(2)$ of the density $N$$\times$$N$ matrix $\rho$ onto
density matrices $\rho(1)$ and $\rho(2)$ following a particular
rule.

Let $N=n m$; this means that we can consider the $N$$\times$$N$
matrix $\rho$ as an analog of the $N$$\times$$N$ matrix $\rho(1,2)$
of bipartite system. We write a new inequality in view of the rule
we found.

First, we present the matrix $\rho$ in the block form $~\rho=
\left(\begin{array}{cccc}
\rho_{11}&\rho_{12}&\cdots&\rho_{1n}\\
\rho_{21}&\rho_{22}&\cdots&\rho_{2n}\\
\cdots &\cdots &\cdots&\cdots\\
\rho_{n1}&\rho_{n2}&\cdots&\rho_{nn}\end{array}\right).$\\[2mm]
\noindent
The blocks $\rho_{kj} (k,j=1,2,\ldots,n)$ in this matrix are the
 $m$$\times$$m$-matrices.\\[1mm] The density $n$$\times$$n$-matrix
 $\rho(1)$ turns out to be
$~~\rho(1)=\left(\begin{array}{cccc}
\mbox{Tr}\,\rho_{11}&\mbox{Tr}\,\rho_{12}&\cdots&\mbox{Tr}\,\rho_{1n}\\
\mbox{Tr}\,\rho_{21}&\mbox{Tr}\,\rho_{22}&\cdots&\mbox{Tr}\,\rho_{2n}\\
\cdots &\cdots &\cdots&\cdots\\
\mbox{Tr}\,\rho_{n1}&\mbox{Tr}\,\rho_{n2}&\cdots&\mbox{Tr}\,\rho_{nn}\end{array}\right).$\\[2mm]
The density $m$$\times$$m$ matrix $\rho(2)$ turns out to be the sum
of $n$ blocks $\rho_{kk}$, i.e.,
$~\rho(2)=\sum_{k=1}^n\rho_{kk}.$ %\\[2mm]

The matrix $\rho$ in our consideration can be the density matrix of
any state, e.g., the density matrix of a single qudit state. The
matrices $\rho(1)$ and $\rho(2)$ are obtained from this matrix by
the same rule as the density matrices of the subsystems in the
bipartite system are obtained using the standard partial trace
procedure. In view of this fact, the matrix inequality corresponding
to the subadditivity condition holds:
\begin{eqnarray*} %\label{LM6}
&&~~-\mbox{Tr}\left\{\left(\begin{array}{cccc}
\rho_{11}&\rho_{12}&\cdots&\rho_{1n}\\
\rho_{21}&\rho_{22}&\cdots&\rho_{2n}\\
\cdots &\cdots &\cdots&\cdots\\
\rho_{n1}&\rho_{n2}&\cdots&\rho_{nn}\end{array}\right)\ln\left(\begin{array}{cccc}
\rho_{11}&\rho_{12}&\cdots&\rho_{1n}\\
\rho_{21}&\rho_{22}&\cdots&\rho_{2n}\\
\cdots &\cdots &\cdots&\cdots\\
\rho_{n1}&\rho_{n2}&\cdots&\rho_{nn}\end{array}\right)\right\}\nonumber\\
&&\leq -\mbox{Tr}\left\{\left(\begin{array}{cccc}
\mbox{Tr}\,\rho_{11}&\mbox{Tr}\,\rho_{12}&\cdots&\mbox{Tr}\,\rho_{1n}\\
\mbox{Tr}\,\rho_{21}&\mbox{Tr}\,\rho_{22}&\cdots&\mbox{Tr}\,\rho_{2n}\\
\cdots &\cdots &\cdots&\cdots\\
\mbox{Tr}\,\rho_{n1}&\mbox{Tr}\,\rho_{n2}&\cdots&\mbox{Tr}\,\rho_{nn}\end{array}\right)\ln
\left(\begin{array}{cccc}
\mbox{Tr}\,\rho_{11}&\mbox{Tr}\,\rho_{12}&\cdots&\mbox{Tr}\,\rho_{1n}\\
\mbox{Tr}\,\rho_{21}&\mbox{Tr}\,\rho_{22}&\cdots&\mbox{Tr}\,\rho_{2n}\\
\cdots &\cdots &\cdots&\cdots\\
\mbox{Tr}\,\rho_{n1}&\mbox{Tr}\,\rho_{n2}&\cdots&\mbox{Tr}\,\rho_{nn}\end{array}\right)\right\}\nonumber\\
&&~~-\mbox{Tr}\left\{(\rho_{11}+\rho_{22}+\cdots
+\rho_{nn})\ln(\rho_{11}+\rho_{22}+\cdots +\rho_{nn})\right\}.
\end{eqnarray*}
This inequality is valid for any Hermitian $N$$\times$$N$-matrix
$\rho$, where $N=nm$, satisfying the condition Tr$\,\rho=1$ and the
nonnegativity condition of the eigenvalues of this matrix. If the
number $N=nm=\widetilde n\widetilde m$, the same matrix $\rho$
satisfies two different inequalities, since one can organize the
block structure of the matrix $\rho$ differently.

\section{Conclusions}
\pst
To conclude, we point out the main results of our study.

We obtained a new inequality for an arbitrary nonnegative Hermitian
matrix with unit trace. If one interprets this matrix as the density
matrix of a two-qubit system, this mathematical inequality is just
the Bell--CHSH inequality with bound~2 for separable states or
Tsirelson bound $2\sqrt 2$ for entangled states. The inequality is
also valid for the density matrix of indivisible systems like the
four-level atom. The latter system can be realized as an artificial
atom in the superconducting circuit based on Josephson-junction
technique~\cite{OlgaDodon1990,ZeilingerJPA,ZeilingerPRB,Kiktenko1,Kiktenko2,GlushkovJRLR}.
The inequalities for such physical systems characterize correlations
like the Bell correlations. For indivisible systems, we call these
correlations {\it the hidden Bell correlations}.

Our statement is as follows.

All kinds of quantum correlations associated either with
entanglement~\cite{Schrod30} or the steering phenomenon available in
multipartite systems are also available in indivisible systems. Bell
correlations for the two-qubit system~\cite{Quan-arXiv} can be found
as hidden Bell correlations in indivisible systems as well.

The examples of qudit with $j=3/2$ or systems like qutrit with the
density 3$\times$3 matrix, which can be considered as a
4$\times$4-matrix with some zero matrix elements, demonstrate the
validity of the inequality obtained.

\section*{Acknowledgments}
\pst We thank the Organizers of the 46th Winter Colloquium on the Physics of
Quantum Electronics (PQE-2016, January 3--8, 2016, Snowbird, Utah,
USA) and especially Prof.~Marlan~O.~Scully for invitation and kind
hospitality.

\section*{Appendix. Inequality for the Hermitian 4$\times$4 Matrix}
\pst Given the 4$\times$4 matrix $f$ such that $f=f^\dagger$.
Let us introduce a 4$\times$4 matrix $\rho(x)$ such that
$$
\rho(x)=(4x+\mbox{Tr}\,f)^{-1}(f+x1),$$
where 1 is the unity 4$\times$4 matrix and $x$ is a real number. If
$f_j$ $(j=1,2,3,4)$ are eigenvalues of the matrix $f$, the matrix
$\rho(x)=\rho^\dagger(x)$, Tr$\,\rho(x)=1$, and $\rho(x)\geq 0$ for
$x>|f_j|$.

Now we construct the stochastic 4$\times$4 matrix
$\Omega_{\alpha\beta}$ with matrix elements
$$
\Omega_{\alpha\beta}(x)=\left(U^{(\alpha)}\rho(x)
U^{(\alpha)\dagger}\right)_{\beta\beta},
$$
where $U^{(\alpha)}$ are unitary 4$\times$4 matrices of the form
$$U^{(1)}=u_1\times u_3,\qquad U^{(2)}=u_1\times u_4,
\qquad U^{(3)}=u_2\times u_3,\qquad U^{(4)}=u_2\times u_4,$$ with
$u_1$, $u_2$, $u_3$, and $u_4$ being arbitrary unitary 2$\times$2
matrices. Then, for $x>|f_j|$, the following inequality takes place
$$
\left|\sum_{\alpha,\beta=1}^4I_{\beta\alpha}\Omega_{\alpha\beta}(x)
\right|\leq 2\sqrt 2.$$ If $f_j>0$, one has the inequality
$$
\left|\sum_{\alpha,\beta=1}^4I_{\beta\alpha}\left(U^{(\alpha)}f
U^{(\alpha)\dagger}\right)_{\beta\beta}\right|\leq 2\sqrt
2(\mbox{Tr}\, f),$$ where the matrix
$I_{\beta\alpha}=\left(\begin{array}{cccc}1&-1&-1&1\\1&-1&-1&1\\1&-1&-1&1\\-1&1&1&-1
\end{array}\right).$ If Tr$\,f=1$, $f_j>0$, and the 4$\times$4
matrix $f$ has the separable form, i.e.,
$$
f=\sum_sp_sf_1^{(s)}\times f_2^{(s)},\qquad 1\geq p_s\geq 0,\qquad
\sum_sp_s=1,$$
where the 2$\times$2 matrices $f_1^{(s)}$ and $f_2^{(s)}$ are
nonnegative Hermitian matrices with unit trace, one has the
inequality
$$
\left|\sum_{\alpha,\beta=1}^4I_{\beta\alpha}\left(U^{(\alpha)}f
U^{(\alpha)\dagger}\right)_{\beta\beta}\right|\leq 2.$$ In the
quantum information context, the matrix inequalities obtained can be
interpreted as the inequa\-lities for quantum observables $f$
characterizing hidden correlations in both composite and
noncomposite systems. Also if the matrix $f$ can be considered as
the density matrix $(f=\rho)$ of the four-level atom or qudit with
$j=3/2$, the inequalities reflect the presence of hidden Bell
correlations in the noncomposite (indivisible) systems.

\end{document}